\documentclass[12pt]{article}
\usepackage{amsmath}
\usepackage{amssymb}
\usepackage{times}

\newcommand{\Gr}{\mathcal{G}\textit{r}}

\newcommand{\ZH}{\mathcal{ZH}}
\newcommand{\V}{\mathcal{V}}
\newcommand{\Hh}{\mathcal{H}}
\newcommand{\HH}{\mathbb{H}}
\newcommand{\U}{\mathcal{U}}     
\newcommand{\W}{\mathcal{W}}      
\newcommand{\C}{\mathbb{C}} 
\newcommand{\CR}{\mathcal{CR}} 
\newcommand{\R}{\mathbb{R}} 
\newcommand{\Oo}{\mathcal{O}} 
\newcommand{\F}{\mathbb{F}} 
\newcommand{\Spin}{\mathcal{S}\textit{pin}}
\newcommand{\SU}{\mathcal{SU}}
\newcommand{\SL}{\mathcal{SL}}
\newcommand{\SO}{\mathcal{SO}}
\newcommand{\T}{\mathcal{T}} 
\newcommand{\Q}{\mathcal{Q}} 
\newcommand{\AQ}{\mathcal{AQ}} 
\newcommand{\GL}{\mathcal{GL}}
\newcommand{\PV}{\mathcal{PV}}
\newcommand{\PX}{\mathcal{PX}}
\newcommand{\X}{\mathcal{X}}
\newcommand{\PT}{\mathcal{PT}}
\newcommand{\PN}{\mathcal{PN}}
\newcommand{\N}{\mathcal{N}}

\begin{document}
\thispagestyle{empty}
\vspace*{30pt}\begin{center}
{\LARGE\textbf{Twistor theory and the four-dimensional Quantum Hall effect of Zhang and Hu}}\bigskip\vspace{10pt}\\
\textmd{George Sparling\\Laboratory of
Axiomatics\\Department of Mathematics\\ University of Pittsburgh\\Pittsburgh,
Pennsylvania, USA}\vspace{10pt}\\
\vspace{50pt}
{\large\textbf{Abstract}}\\\vspace{5pt}
\begin{quote}\textbf{
The construction by Zhang and Hu of a four-dimensional analogue of the Quantum Hall effect is generalized and recast as a purely geometrical
theory, using the languages of Lie group theory and twistor theory. It emerges that the Zhang-Hu quantum liquid lies
naturally in twistor space and is apparently more primitive than space-time itself, in accordance with twistor philosophy.  The quantum liquid is
then the glue that holds space-time together.  It is argued that the theory
is inherently chiral and time asymmetric, consonant with previous conjectures. }
\end{quote}
\end{center}
\subsection*{Introduction}
Recently  Zhang and Hu presented a groundbreaking extension of the theory of the quantum Hall effect from two to four dimensions
\cite{A1}-\cite{A121}. The purpose of the present work is two-fold: to present a substantial simplification and generalization of some of their work
and to discuss its relation to twistor theory \cite{A14}-\cite{A20}.  The main conclusion is that, in two and four dimensions, the quantum Hall effect
may be cast in purely geometrical language, the geometry being that of spinor and twistor theory.   The relevant symmetry  group is
$\U(n, \HH)$ the group of all
$n\times n$ quaternionic unitary matrices: a real Lie group of real dimension $n(2n+1)$ \cite{A13}.  \begin{itemize}\item  For $n = 1$, the group
is isomorphic to the spin group in three real Euclidean dimensions, leading to the two-dimension theory of the quantum Hall effect, in the style of
Haldane \cite{A6}.
\item  For
$n =2$ the group is isomorphic to the spin group in five real Euclidean dimensions, leading to the Zhang-Hu theory of the four dimensional quantum
Hall effect \cite{A1}-\cite{A2}.  \item  For $n > 2$, the geometrical interpretation is presently unclear, although the $n = 4$ case could  relate
to the analysis of quantum ensembles of four dimensional universes (see below).  In this interpretation, the Lie algebra of $\U(n, \HH)$ is
extended to a graded Lie algebra (also called a superalgebra).\end{itemize} In the following, we first discuss the use of Grassman variables to
analyze the Pfaffian matrix of a symplectic form.  Then we show that the Zhang-Hu state representing at each level
$r$ the highest exterior product of the orthonormal states for that level is expressible invariantly as a simple Pfaffian matrix, provided that
$r$ is odd.  To show that our version of this state is non-zero, we develop the appropriate Hilbert space theory and calculate the norm of the
state directly: this norm turns out to be non-zero, so we are done.
\\\\ Finally, we analyze the so-called edge-states.  We show that the "edge" in question is the standard $(2,2)$ hyperquadric in twistor space
\cite{A14}-\cite{A20}. This is a quadratic $\CR$ submanifold of $\C^4$ with symmetry group the pseudo-unitary group $\U(2,2, \C)$ \cite{A21}-\cite{A23}.
Since this group maps directly to the conformal group of real (four-dimensional) Minkowski spacetime, the structure of the hyperquadric seems to lead
directly to relativity.  At the particle level, it appears that the edge-states give rise to deformations of the $\CR$ structure of the hyperquadric (or
of appropriate $\CR$ bundles over the hyperquadric).  Such deformations are parametrized by
$\CR$ first cohomology groups \cite{A23}.  These groups, in turn, are well-known to correspond to massless particles \cite{A19}.  Their spectrum is
naturally that of one each of all possible helicities, exactly the spectrum found by Zhang and Hu. We discuss various possible limiting procedures that
might be used with these edge-states and outline a possible connection with the author's previous proposals linking twistor theory, time asymmetry, string
theory and the Grothendieck theory of 'dessins d'enfants" \cite{A24}-\cite{A267}.      
\eject
\subsection*{Grassman variables}
Let $\W$ be a vector space of finite dimension $n$, over a field $\F$ of characteristic zero; denote by $\W^*$, the dual space of $\W$.  Associated
to
$\W$ and $\W^*$ are the tensor algebra based on $\W$ and $\W^*$ and the exterior algebras of $\W$ and $\W^*$: the latter are written
$\Omega(\W)$ and $\Omega(\W^*)$, respectively. We use the convention that for each positive integer $k$, the exterior product $\Pi_{i =1}^k w_i
= w_1w_2
\dots w_k$ of  any
$k$ vectors $w_1, w_2, \dots, w_k$ from
$\W$ is given in terms of their tensor product by the formula: 
\begin{equation} w_1w_2\dots w_k = \frac{1}{k!}\sum_{\tau\in
\mathcal{S}_k}\textrm{sgn}(\tau)w_{\tau(1)}\otimes w_{\tau_2}\otimes\dots\otimes w_{\tau(k)}.\end{equation}
Here $\mathcal{S}_k$ is the permutation group on $k$ elements and $\textrm{sgn}(\tau)$ is the sign of $\tau$.
\\\\Put $\Omega^0(\W) = \F$, and for each positive integer $k$, denote by $\Omega^k(\W)$ the vector subspace of
$\Omega(W)$ spanned by all exterior products of $k$-vectors taken from $\W$ (in particular $\Omega^1(\W) = \W$).  Also put
$\Omega^k(\W) = \{0\}$, for $k< 0$. The dimension of $\Omega^k(\W)$ is $\binom{n}{k}$ for $0\le k\le n$ and is zero otherwise.  In particular
$\Omega^n(\W)$ is a one-dimensional vector space. Any non-zero element of $\Omega^n(\W)$ is called an orientation for $\W$.  The spaces
$\Omega^n(W)$ and $\Omega^n(\W^*)$ are canonically dual, so an orientation for $\W$ gives one for $\W^*$ and vice-versa.
\\\\We consider (odd)
Grassman variables taking values in $\W$.  If
$\theta$ is such a variable, then the polynomials in
$\theta$ form an algebra naturally isomorphic to $\Omega(\W^*)$. If $\theta$ and $\phi$ are two such variables, it is
understood that they mutually anti-commute.  
\\\\Associated to the variable
$\theta$ is a natural odd derivation, denoted
$\delta_\theta$ which takes values in $\W^*$ and which obeys the normalization condition:
\begin{equation}\delta_\theta \otimes \theta = \delta.\end{equation}
Here $\delta\in \W^*\otimes \W$ is the Kronecker delta tensor. The operator $\delta_\theta$ obeys a version of Taylor's theorem: acting on any
polynomial $f(\theta)$, we have:
\begin{equation} \exp(\lambda.\delta_\theta)f(\theta) = f(\theta + \lambda), \hspace{10pt} \exp(\theta.A.\delta_\theta)f(\theta) =
f(\exp(A)\theta).\end{equation} Here $\lambda$ is any $\W$-valued (odd) Grassman variable independent of $\theta$, $f$ is any polynomial in $\theta$
and $A$ is any (even) element of
$\W\otimes \W^*$ independent of $\theta$.
\\
\\
If $\theta_i, i = 1, 2, \dots, k$ are Grassman variables, we write $\int_\theta$ for the operator $\prod_{j = 1}^k \delta_{\theta_{k-j}}$.  This
extracts the coefficient of
$\prod_{i =1}^k \theta_i$ from a polynomial in the $\theta_j$. \eject\noindent
\subsection*{Symplectic forms, invariant Pfaffians and Determinants}
Let $\V$ be a vector space over the complex field, of complex dimension $2n$, with $n$ a positive integer.  Let $\omega$ be a symplectic form on
$\V$: a non-degenerate skew bilinear form on $\V$, also regarded as an element of $\Omega^2(\V^*)$. Given $\omega$, a symplectic basis relative to
$\omega$ is a a basis
$\{e_i, i = 1,2,
\dots, 2n\}$ of $\V^*$, such that $\omega$ has the decomposition:
\begin{equation} \omega  = 2\sum_{i = 1}^n e_{2i - 1} e_{2i} =  \sum_{i =1}^n (e_{2i-1}\otimes e_{2i} -
e_{2i}\otimes e_{2i-1}).\end{equation}
Such a symplectic basis always exists.
\begin{itemize}\item 
Let $\theta$ and $\phi$ be Grassman variables, taking values in $\V$.
Then we have the Grassman symplectic identity:
\begin{equation} c_n\omega(\theta, \theta)^n\omega(\phi, \phi)^n = \omega(\theta, \phi)^{2n}, \hspace{20pt}c_n =
(-\frac{1}{4})^n\binom{2n}{n}.\end{equation} Proof:\\
For a symplectic basis $\{e_i, i = 1,2, \dots, 2n\}$ of $V^*$, relative to $\omega$, we put $\theta_i = e_i(\theta)$ and
$\phi_i = e_i(\phi)$, for $i = 1, 2, \dots 2n$. Then we have:
\begin{equation} \omega(\theta,
\theta) =2
\sum_{k = 1}^n \theta_{2k-1}\theta_{2k},\end{equation}\begin{equation} \omega(\phi,
\phi) =2
\sum_{k = 1}^n \phi_{2k-1}\phi_{2k}, \end{equation}\begin{equation}  \omega(\theta,
\phi) =
\sum_{k = 1}^n (\theta_{2k-1}\phi_{2k}+\phi_{2k-1}\theta_{2k}).  \end{equation}
Then the desired equation is checked by direct expansion.  Both sides of the equation, when expanded, give the quantity $(2n)!\prod_{j =
1}^{2n}\theta_j\phi_j$.
\end{itemize}
Define a skew multilinear form $\textrm{Pf}(\omega)\in \Omega^{2n}(\V^*)$ on $\V$, in $2n$ arguments, by the formula:
\begin{equation} \textrm{Pf}(\omega)(\theta, \theta, \dots, \theta) = \frac{1}{n!2^n}\omega(\theta, \theta)^n.\end{equation}
Then $\textrm{Pf}(\omega)$, called the Pfaffian of $\omega$, is an orientation for $\V^*$.
With respect to a symplectic basis for $\omega$, we have:
\begin{equation}  \textrm{Pf}(\omega) = e_1 e_2\dots  e_{2n},\end{equation}
\begin{equation}  \textrm{Pf}(\omega)(\theta, \theta, \dots, \theta) = \prod_{j = 1}^{2n} \theta_j.\end{equation}
Similarly, the quantity $\omega(\theta, \phi)^{2n}$ defines an element $\textrm{Det}(\omega)$ of $\Omega^{2n}(\V^*)\otimes \Omega^{2n}(\V^*)$
called the determinant of $\omega$, by the formula:
\begin{equation}  \textrm{Det}(\omega)(\theta, \theta, \dots, \theta, \phi, \phi, \dots, \phi) = \frac{(-1)^n}{(2n)!}\omega(\theta,
\phi)^{2n}.\end{equation} With respect to the symplectic basis considered above, we have:
\begin{equation} \textrm{Det}(\omega) = e_1 e_2\dots  e_{2n}\otimes  e_1 e_2\dots  e_{2n}.\end{equation}
Then, with these definitions, the Grassman symplectic identity becomes the following:
\begin{equation} \textrm{Pf}(\omega)\otimes \textrm{Pf}(\omega) = \textrm{Det}(\omega).\end{equation} 
Note that if $\omega = \sum_{1\le i, j\le 2n} \omega_{ij}\epsilon_i\epsilon_j$, where $\{\epsilon_i, i = 1,2, \dots, 2n\}$ is any basis of
$\Omega^1(V^*)$, and where the coefficient matrix $\omega_{ij}$ is skew: $\omega_{ij} = - \omega_{ji}$, then we have
$\int_\epsilon \textrm{Pf}(\omega) = \textrm{pf}(w)$ and $(\int_\epsilon\otimes \int_\epsilon)\textrm{Det}(\omega) = \det(\omega)$, where
$\textrm{pf}(\omega)$ and
$\det(\omega)$ are the ordinary Pfaffian and determinant of the $2n$ by $2n$ matrix $\omega_{ij}$, respectively. Now
$\textrm{pf}(\omega)$ and $\det(\omega)$ are numbers and the Grassman symplectic identity becomes the following:
\begin{equation} \textrm{pf}(\omega)^2 = \det(\omega).\end{equation}
Explicitly, we have $n!2^n\textrm{pf}(\omega) = \sum_{\tau\in \mathcal{S}_{2n}}\prod_{j =1}^n \textrm{sgn}(\tau)\omega_{\tau_{2j-1}, \tau_{2j}}$.  
For the present work a key feature of $\textrm{pf}(\omega)$ is that if, for $1\le p \ne q\le 2n$, the $p$-th  and $q$-th rows of the matrix
$\omega_{ij}$  are interchanged and at the same time the
$p$-th and
$q$-th columns are interchanged, to preserve skewness, then $\det(\omega)$ stays unchanged, whereas $\textrm{pf}(\omega)$ changes sign.
\subsection*{Quaternionic unitary and pseudo-unitary structures}
Let $n$ be a positive integer and let $p$ and $q$ be non-negative integers, such that $p + q = n$.  Denote by $I_{p,q}$ a diagonal $n\times
n$ matrix with the first $p$ entries along the diagonal being $1$ and the last $q$ entries $-1$.  The quaternion pseudo-unitary
group
$\U(p,q, \HH)$ is the group of all
$n\times n$ matrices, $M$, with quaternionic entries, obeying $M^*I_{p,q}M = I_{p,q}$, where $M^*$ is the transpose quaternion conjugate matrix of
$M$.  Then
$\U(p,q,\HH)$ (isomorphic to $\U(q,p, \HH)$) is a real Lie group of dimension
$n(2n+1)$ and is non-compact unless $p$ or $q$ vanishes.  The compact group $\U(n, 0, \HH)$ is written $\U(n, \HH)$ and is called the
quaternionic unitary group. We have the isomorphisms:
\begin{itemize}\item $\U(1,\HH) = \SU(2,\C) = \Spin(3,\R)$,
\item $\U(2, \HH) = \Spin(5,\R)$; \hspace{10pt} $\U(1,1, \HH) = \Spin(1,4,\R)$.
\end{itemize}
Let $\V$ be a complex vector space of complex dimension $2n$.  Then $\V$ is said to have a quaternionic pseudo-unitary
structure if $\V$ is equipped with the following:
\begin{itemize}\item A quaternionic conjugation: for any $v\in \V$ its conjugate is denoted $\overline{v}$.  The conjugation is complex antilinear
and obeys $\overline{\overline{v}} = - v$, for any $v \in \V$. In particular $\V$ has no real (self-conjugate) elements.  The conjugation extends
naturally to the tensor algebra of $\V$ and $\V^*$, such that $\overline{A\otimes B} = \overline{A}\otimes \overline{B}$ for any tensors $A$ and
$B$ and $\overline{\alpha}.v = - \overline{\alpha.\overline{v}}$, for any $v\in \V$ and any $\alpha\in \V^*$.
\item A complex bilinear symplectic form, $\omega$, that is self-conjugate: $\overline{\omega} =
\omega$.
\end{itemize}
Given a quaternionic pseudo-unitary structure for $\V$, there is a natural hermitian sesquilinear form $g$ on
$\V$, given by the formula, valid for any
$v$ and
$w$ in $\V$:
\begin{itemize}\item $g(v, w) = \omega(v, \overline{w}) = \overline{\omega(w, \overline{v})} = \overline{g(w,v)}$.
\end{itemize}
Note that $g$ is complex linear in its first argument and antilinear in its second. The quaternionic pseudo-unitary structure is of type
$(p,q)$, with structure group $\U(p, q, \HH)$, if the hermitian form
$g$ is of type
$(p,q)$; the structure is unitary, with structure group $\U(n, \HH)$ if
$g$ is definite. Henceforth we restrict to the quaternionic unitary case and require that
$g$ be positive definite: for each $v\ne 0$ in $\V$, we have $\omega(v,\overline{v}) > 0$.  Also define
$|v| =
\sqrt{\omega(v,\overline{v})}$, for any
$v
\in \V$.
\begin{itemize}
\item A normalized basis for the quaternionic unitary vector space $\V$ is a symplectic basis such that $e_{2i} = \overline{e_{2i-1}}$ for $i =
1,2,
\dots n$ and for any
$v
\in \V$,
$|v|^2 =
\sum_{k = 1}^{2n} |e_i(v)|^2$. 
\end{itemize}
A normalized basis always exists for any quaternion unitary vector space $\V$.
\eject\noindent\subsection*{The Grassman Zhang-Hu state}
For $\V$ a complex vector space of complex dimension $2n$ and for any positive integer $r$, denote by $\V^r$ the $r$-fold symmetric tensor product
of
$\V$ with itself: in particular
$\V^1 = \V$;  also, for each $v\in \V$, denote by $v^r\in \V^r$, the $r$-fold symmetric tensor product of $v$ with itself.  The complex dimension
of $\V^r$ is given by the binomial coefficient
$\binom{r + 2n - 1}{2n-1}$.
\begin{itemize}\item For $r$ an odd positive integer and for $n$ a positive integer, the binomial coefficient
$\binom{r+ 2n-1}{2n-1}$ is even. Consequently
$\V^r$ has even dimension, provided that $r$ is odd.\\\\ 
Proof: \\Put $r = 2s + 1$, with $s$ a non-negative integer.  Then we have the following integer identity, which gives us the required result
immediately:
\begin{equation}  \binom{2n\hspace{-2pt}+\hspace{-2pt}r\hspace{-2pt}-\hspace{-2pt}1}{r} =
\binom{2n\hspace{-2pt}+\hspace{-2pt}2s}{2s\hspace{-2pt}+\hspace{-2pt}1} = 2\frac{(n + s)\binom{n+s - 1}{s}\prod_{m = n}^{n+s -1}(2m + 1)}{\prod_{k = 0}^s
(2k+1)}.\end{equation}
\end{itemize}
Henceforth $r$ denotes an odd positive integer and $\V$ a complex vector space of dimension $2n$ with a quaternionic unitary structure. Then $\V^r$
inherits from
$\V$ a natural quaternionic unitary structure. Denote the symplectic form of $\V^r$ by $\omega_r$, so that $\omega_1 = \omega$.  The conjugation
and the symplectic form
$\omega_r$ are normalized by the formulas, valid for any $v$ and $w$ in $\V$:
\begin{equation}  \overline{v^r} = (\overline{v})^r, \hspace{10pt} \omega_r(v^r, w^r)  = (\omega(v,w))^r.\end{equation}
\begin{itemize}\item 
Let $r$ be a fixed odd positive integer.  Let $\theta$ be a Grassman variable taking values in $\V^r$.  Put $M = \frac{1}{2}\binom{r + 2n -
1}{2n-1}$.  Then the Grassman Zhang-Hu state $\ZH(\theta)$ is, by definition:
\begin{equation}  \ZH(\theta) = \frac{1}{M!2^M}\omega_r(\theta, \theta)^M = Pf(\omega_r)(\theta, \theta, \dots, \theta).\end{equation}
\item Let $\delta$ be the $(\V^*)^r$ valued derivation dual to $\theta$, so $(v^r.\delta)\theta = v^r$, for any $v\in \V$.  Let $\lambda_i$, for
$1\le i
\le 2M$ be Grassman variables. Then we may extract from the Grassman Zhang-Hu state a multivariate polynomial, denoted $\ZH(v_1, v_2, \dots,
v_{2M})$, with
$2M$ arguments taken from $\V$, as follows:
\begin{equation}\ZH(v_1, v_2, \dots v_{2M}) = \int_{\lambda}\frac{1}{(2M)!}(\sum_{i = 1}^{2M}\lambda_i
v_i^r.\delta)^{2M}\ZH(\theta).\end{equation}
\end{itemize}
We can rewrite this expression, using the Grassman Taylor theorem, in various ways:
\begin{equation} \ZH(v_1, v_2, \dots v_{2M}) = (-1)^M\left(\prod_{i = 1}^{2M}v_i^r.\delta\right) \ZH(\theta)\end{equation}\begin{equation} = \int_\lambda
\frac{1}{M!(2M)!2^M}(\sum_{i = 1}^{2M}\lambda_i v_i^r.\delta)^{2M}\omega_r(\theta,\theta)^M \end{equation}\begin{equation} = \frac{1}{2^MM!}\int_\lambda
[\exp(\sum_{i = 1}^{2M}\lambda_i v_i^r.\delta))\omega_r(\theta,\theta)^M]_{\theta = 0} \end{equation}\begin{equation} = \frac{1}{2^MM!}\int_\lambda
[(\omega_r(\theta + \sum_{i = 1}^{2M}\lambda_i v_i^r,\theta +  \sum_{i = 1}^{2M}\lambda_i
v_i^r))^M]_{\theta = 0}\end{equation}\begin{equation} = \frac{1}{2^MM!}\int_\lambda (\omega_r(\sum_{i = 1}^{2M}\lambda_i
v_i^r,\sum_{i = 1}^{2M}\lambda_i
v_i^r))^M\end{equation}\begin{equation} = \frac{1}{2^MM!}\int_\lambda (\sum_{1\le i, j\le 2M}\lambda_i\lambda_j\omega_r(v_i^r,
v_j^r))^{M}\end{equation}\begin{equation} = \textrm{pf}(\Omega_r), \hspace{10pt}\end{equation}\begin{equation}(\Omega_r)_{ij} = \omega_r(v_i^r, v_j^r) =
\omega(v_i, v_j)^r.\end{equation} Note that since $r$ is odd, $\Omega_r$ is a skew matrix and the Pfaffian of $\Omega_r$ is alternating in the
variables $\{v_i^r, i = 1, 2, \dots, 2M\}$.  So provided that the Pfaffian is non-zero in general, $\ZH$ must agree up to a non-zero scale factor with
the state constructed by Zhang and Hu.  We will establish that the state is non-zero by introducing an appropriate norm and showing that the norm of the
state $\ZH$ is non-zero.
\subsection*{A Hilbert space of entire functions and the norm of the $\ZH$ state}
Consider the space of entire functions on $\V$, $\Oo(\V)$, where $\V$ is a quaternionic unitary complex vector space of complex dimension $2n$,
with self-conjugate symplectic form
$\omega$.  We give $\Oo(\V)$ a hermitian inner product $G(f, g)$, defined for suitable entire $f$ and
$g$ on $\V$, by the formula:
\begin{equation} G(f, g) = \int_{\V}  f(z)\overline{g}(\overline{z})e^{-\omega(z, \overline{z}) + \frac{i}{2\pi}\omega(dz,
d\overline{z})}.\end{equation} Here the integration is of the top form, taken over all the real $4n$-dimensional space $\V$. The
orientation is chosen so that
$G(1,1) > 0$.  We put $|f| =\sqrt{ G(f,f)}$.  The Hilbert space
$\Hh$ is then the space of entire functions $f$, on $\V$, with $|f|$ finite.  Define $\Hh_+$ and $\Hh_-$ to be
the Hilbert subspaces, of $\Hh$, consisting of even and odd entire functions, respectively.  Also define $\Hh_s$, for any non-negative integer $s$, 
as the subspace of all polynomials homogeneous of degree $s$, with the induced Hilbert space structure; then $\Hh_s$ lies in  $\Hh_+$ iff $s$ is even and
in
$\Hh_-$ iff
$s$ is odd.   The inner product of
$\Hh$ has many nice properties, including the following, in which
$a$ and
$b$ are any elements of
$\V^*$ and
$s$ and $t$ are non-negative integers:
\begin{itemize}\item The orientation condition: $G(1,1) = 1$.
\item $G(e^{a.z}, e^{b.z}) = e^{\omega^{-1}(a,\overline{b})}$.
\item $G((a.z)^s,(b.z)^t) = \delta_{st} s!\omega^{-1}(a,\overline{b})^s$ and $|(a.z)^s|^2 = s!(\omega^{-1}(a,\overline{a}))^s$.
\item $\Hh_+$ and $\Hh_-$ are orthogonal; $\Hh_s$ and $\Hh_t$ are orthogonal whenever $s \ne t$.
\item For any $f$, such that $f(z)$ and the $\V^*$-valued derivative $\partial_z f(z)$ are normalizable, then $(a.z)f(z)$ is
normalizable and we have: 
\begin{equation} |(a.z)f|^2 = |a|^2|f|^2 + |\omega^{-1}(\partial_z(f), \overline{a})|^2.\end{equation}
\item For any normalizable entire function $f$, we have:
\begin{equation}|f|^2 = [\overline{f}(\omega^{-1}(\partial_z)) f(z)]_{z = 0}.\end{equation}
\item Put $(T_{x}f)(z) = f(z - x)$, for any $x\in \V$ and any $z \in \V$.  Then we have that $f$ is normalizable iff $T_{x}(f)$ is
normalizable, for any $x \in \V$ and we have:
\begin{equation} |e^{a.z} f|^2 = e^{\omega^{-1}(a,\overline{a})}|T_{\omega^{-1}(\overline{a})}(f)|^2.\end{equation}
\end{itemize}
In particular, if $e_i, i = 1,2,..,2n$ is an orthonormal frame for $\V^*$: $\omega^{-1}(e_i, \overline{e_j}) = \delta_{ij}$,
put
$z_i = e_i(z)$. Then the following monomial entire functions on $\V$ form an orthonormal frame for $\Hh$:
\begin{equation} (r_1, r_2, \dots, r_{2n}) = \prod_{j = 1}^{2n} \frac{z_j^{r_j}}{\sqrt{(r_j)!}}.\end{equation}
If we restrict to the case that $\sum_{j = 1}^{2n} r_j = r$, and to the cases that $n = 1$ or $2$, the monomials give an orthonormal frame for
$\Hh_r$, the same as that provided by Haldane, when $n = 1$ and by Zhang and Hu when $n  = 2$.  \\
\\
Now we may calculate the norm of the $\ZH$ state as follows:
\begin{equation}M!^24^{M}|\ZH(v_1, v_2, \dots, v_{2M}|^2 = |(\prod_{i = 1}^{2M}v_i^r.\delta) \omega_r(\theta, \theta)^M|^2\end{equation}
\begin{equation} = (r!)^{2M}(\omega_r^{-1}(\delta, \overline{\delta}))^{2M}(\omega_r(\theta, \theta)^M\omega_r(\overline{\theta},
\overline{\theta})^M)\end{equation}
\begin{equation} = (r!)^{2M}(2M)!\exp(\omega_r^{-1}(\delta, \overline{\delta}))(\omega_r(\theta, \theta)^M\omega_r(\overline{\theta},
\overline{\theta})^M)|_{\theta = 0}\end{equation}\begin{equation} = (r!)^{2M}(2M)!(\omega_r(\theta - \omega_r^{-1}(\overline{\delta}) ,
\theta - \omega_r^{-1}(\overline{\delta}))^M\omega_r(\overline{\theta},
\overline{\theta})^M)|_{\theta = 0}\end{equation}\begin{equation} = (r!)^{2M}(2M)!\omega_r^{-1}(\overline{\delta},
\overline{\delta})^M\omega_r(\overline{\theta},
\overline{\theta})^M\end{equation}
Finally, we calculate $X_k = \omega_r^{-1}(\overline{\delta}, \overline{\delta})^k\omega_r(\overline{\theta},
\overline{\theta})^k$, for $k$ a positive integer, by induction:
\begin{equation} X_k  =  \omega_r^{-1}(\overline{\delta}, \overline{\delta})^{k-1}\omega_r^{-1}(\overline{\delta},
\overline{\delta})\omega_r(\overline{\theta},
\overline{\theta})^k\end{equation}\begin{equation} =  2k\omega_r^{-1}(\overline{\delta},
\overline{\delta})^{k-1}(\overline{\delta}.\overline{\theta})\omega_r(\overline{\theta},
\overline{\theta})^{k-1}\end{equation}\begin{equation} =  2k\omega_r^{-1}(\overline{\delta}, \overline{\delta})^{k-1}(2M -
\overline{\theta}.\overline{\delta})\omega_r(\overline{\theta},
\overline{\theta})^{k-1}\end{equation}\begin{equation} =  4k(M - k+1)X_{k-1} = \frac{ 4^{k}k!M!}{(M - k)!}.\end{equation}
Then we have:
\begin{equation} |\ZH(v_1, v_2, \dots, v_{2M})|^2 = \frac{1}{ M!^24^{M}}(r!)^{2M}(2M)!X_M  = (r!)^{2M}(2M)!.\end{equation}
This is non-zero, so we have recovered the Zhang-Hu state, as required.
\subsection*{First quantization, the Hamiltonian, the spectrum and the multiplicities}
The space $\V$ may be regarded as a classical phase space, with the Poisson brackets $\{a.z, b.z\} = \{a.\overline{z}, b.\overline{z}\} = 0$,
$\{a.z, b.\overline{z}\} = i\omega^{-1}(a,b)$, for any constant vectors $a$ and $b$ in $V^*$.  Then the function
$a.z$ is quantized on the Hilbert space of entire holomorphic functions $\Hh$, as a multiplication operator and the function $b.\overline{z}$ as
the derivative operator
$-\omega^{-1}(b,\partial_z)$. Note that the operators $a.z$ and $\overline{a}.\overline{z}$ are mutually adjoint, with respect to the inner
product of $\Hh$. For any constant
$a$ and $b$ in $\V^*$, the operator commutation relations are as follows:
\begin{equation} [a.z,b.z] = 0, \hspace{10pt} [a.\overline{z},b.\overline{z}] = 0, \hspace{10pt} [a.z, b.\overline{z}] =
-\omega^{-1}(a,b).\end{equation} The quantization of
the operator $j = \omega(z, \overline{z})$ is ambiguous, up to a constant.  Here we prefer to use its symmetrization:
\begin{equation} j \rightarrow \frac{1}{2}(\omega(z, \overline{z}) - \omega(\overline{z}, z))  = \frac{1}{2}(z.\partial_z + \partial_z.z )
= z.\partial_z + n.\end{equation} The Zhang-Hu Hamiltonian operator, $E$, is a combination, $E = H - 2J^2$.  Here $H$ is a Casimir
operator formed from the generators of the $\U(n, \HH)$ symmetry group and $J^2$ is the "squared isospin" Casimir operator. The isospin algebra is the
algebra of the operators
$z.\partial_{\overline{z}}$,
$\overline{z}.\partial_z$ and $j - \overline{j}$, where $\overline{j} = \overline{z}.\partial_{\overline{z}} + n$.  Then we have $4J^2
=2z.\partial_{\overline{z}}\overline{z}.\partial_z + 2\overline{z}.\partial_z z.\partial_{\overline{z}} +(j - \overline{j})^2$. Then (using
abstract tensor indices, for clarity
\cite{A17}) we may write
$H = -
\omega^{ij}\omega^{pq}\Lambda_{ip}\Lambda_{jq}$, where
$\Lambda_{ip} =
\frac{1}{2}(E^k_i\omega_{jk} +  E^k_j\omega_{ik})$.  Here 
$\omega_{ij}$ and $\omega^{ij}$ are the indexed versions of $\omega$ and $\omega^{-1}$, respectively (they are related by the formula
$\omega_{ik}\omega^{jk} =\delta_i^j$); also the
$E_j^i = z^i\partial_j +
\overline{z}^i\overline{\partial}_{j}$ are the generators of the quaternionic linear action on
$\V$. Simplifying
$H$, gives the equation:
\begin{equation} -2H = E_i^pE_j^q\omega_{pq}\omega^{ij} - E_i^jE_j^i\end{equation}\begin{equation} = (z^p\partial_i +
\overline{z}^p\overline{\partial}_i)(z^q\partial_j + \overline{z}^q\overline{\partial}_j)\omega_{pq}\omega^{ij} - (z^j\partial_i +
\overline{z}^j\overline{\partial}_i)(z^i\partial_j +
\overline{z}^i\overline{\partial}_j)\end{equation}\begin{equation} = 2z^p\overline{z}^q\omega_{pq}\omega^{ij}\partial_i\overline{\partial}_j -
2(\overline{z}^j\partial_j)( z^i\overline{\partial}_i) - j^2 - \overline{j}^2 + 2\overline{j} + 2n^2 - 2n.\end{equation} Note
that the two operators
$\omega^{ij}\partial_i\overline{\partial}_j$ and $z^i\overline{\partial}_i$ mutually commute.  If now we operate on a function
$f(z^i, \overline{z}^j)$ which is killed by each of those operators, we get:
\begin{equation} 2H =    j^2 + \overline{j}(\overline{j} - 2) - 2n(n-1), \hspace{10pt}4J^2 = (j - \overline{j})(j - \overline{j} + 2),
\end{equation}\begin{equation} E =    j(\overline{j} -1) - n(n-1).\end{equation}If further $f$ is homogeneous of degrees $p$ in $z^i$ and
$q$ in $\overline{z}^j$, then we have:
\begin{equation} H = \frac{1}{2}(p^2 + q^2 + 2np + 2(n - 1)q),\hspace{10pt}J^2 = \frac{1}{4}(p - q)(p - q + 2), \end{equation}\begin{equation} E = (p +
n)(q + n - 1) - n(n -1).\end{equation} In the cases $n = 1$ and $n = 2$, these formulas exactly agree with the Haldane and Zhang-Hu
energies, respectively.  The corresponding states are polynomials in
$z^i$ and
$\overline{z}^{j}$:
\begin{equation}f_{(p,q)} = F_{i_1i_2\dots i_pj_1j_2\dots j_q}z^{i_1}z^{i_2}\dots z^{i_p}\overline{z}^{j_1}\overline{z}^{j_2}\dots
\overline{z}^{j_q}.\end{equation}Here $p \ge q\ge 0$ and the tensor $ F_{i_1i_2\dots i_pj_1j_2\dots j_q} $ obeys the following conditions:
\begin{equation} F_{i_1i_2\dots i_pj_1j_2\dots j_q}  =  F_{(i_1i_2\dots i_p)(j_1j_2\dots j_q)}, \end{equation}\begin{equation} 
F_{(i_1i_2\dots i_pj_1)j_2\dots j_q}  =  0, \hspace{6pt}\textrm{if}\hspace{6pt}q > 0, \end{equation}\begin{equation} \omega^{i_1j_1} F_{i_1i_2\dots
i_pj_1j_2\dots j_q}  = 0, 
\hspace{6pt}\textrm{if}\hspace{6pt}q > 0.\end{equation} Here, parentheses around tensor indices indicate complete idempotent
symmetrization about the indices contained in the parentheses.  The dimension $d(p, q)$ of the representation is the difference $Y(p,q) - Y(p - 1, q-
1)$, where $Y(p,q)$ is the dimension of a Young's tableau with two rows, one with $p$ boxes the second the
$q$ boxes.  Explicitly, we have:
\begin{equation} Y(p, q) = \frac{(p - q + 1)(2n + p - 1)!(2n + q - 2)!}{(p +  1)!q!(2n -1)!(2n - 2)!},\end{equation}\begin{equation} d(p,
q) = \frac{(p - q+1)(p+q + 2n -1)(p + 2n - 2)!(q + 2n - 3)!}{(2n - 1)!(2n - 3)!(p+1)!q!}.\end{equation} When $n = 2$, we have: $d(p,q)
=\frac{1}{6} (p - q + 1)(p + q + 3)(p + 2)(q + 1)$, again in exact agreement with the Zhang-Hu formula.  In the case $n =1$, a non-zero state necessarily
has $q = 0$.  Then the energy is zero and the multiplicity is $p +1$, in agreement with the Haldane theory.  Henceforth we deal only with the case $q =
0$, where the states are entire functions $f(z^i)$.  Then if $f$ is homogeneous of degree $p$ in $z^i$, the energy is $(n - 1)p$ and the multiplicity is
$d(p, 0) = \binom{p+2n -1}{2n-1}$. 
\subsection*{Second quantization and the graded Lie algebra}
Second quantization enables us to deal with multi-particle states.  Here, if we take the one-particle Hilbert space to be $\Hh$, a  
$k$-particle state may be represented by an entire function $f(z_1, z_2, \dots, z_k)$ of $k$-variables from $\V$.   The state is fermionic in the
sense of Zhang and Hu, iff
$f$ is alternating in \emph{all} its arguments.  The state is at level $r$ if it is homogeneous of degree $r$ in each variable.  In this language
the
$\ZH$ ground state for level $r$ is a non-zero fermionic state at level $r$, of $ \binom{r+2n -1}{2n-1}$ particles and is unique up to a non-zero
scaling.  If such a level $r$ is fixed, and if $s$ is an odd positive divisor of $r$, the particles at levels $\frac{r}{s}$ are said to be
$\frac{1}{s}$-th of a particle at level $r$: the $s$-th power of a fermionic state at level $\frac{r}{s}$ is a state at level $r$.  \\
\\In this
work we would prefer to maintain the usual spin-statistics relation.  For us the twistors are intrinsically spinorial, so entire functions that
are even (those in
$\Hh_+$) should be bosonic in nature, whereas odd entire functions (those in $\Hh_-$) should be fermionic.  Also the multiplication operators $a.z$
and the derivative operators
$b.\partial_z$, for $a \in \V^*$ and $b \in \V$ interchange $\Hh_+$ and $\Hh_-$ so are Fermi-Bose operators.  Consequently the correct
multi-particle Hilbert space (the Fock algebra) for us is the tensor product of the symmetric tensor algebra based on $\Hh_+$ with the exterior
algebra based on
$\Hh_-$.  In particular an entire function
$f(z_1, z_2,
\dots, z_k)$ would only be required to be skew in certain of its arguments iff it is also odd in each of those arguments.  The relevant algebra of
operators is now the graded Lie algebra generated by the odd operators $z$ and $\overline{z}$, which are second quantized as odd derivations of
the Fock algebra.  The graded Lie algebra is then the quotient of the free graded Lie algebra generated by these operators modulo the relations given
by the one-particle \emph{commutation} relations \cite{A33}. Using indices, its even generators are the following operators:
\begin{equation}E^{ij} = z^iz^j,\hspace{10pt} F^{ij} = \overline{z}^i\overline{z}^j, \hspace{10pt}G^{ij} = z^i\overline{z}^j
+\overline{z}^jz^i.\end{equation}   These even generators generate the ordinary quaternionic unitary Lie algebra, $\U(2n, \HH)$, of real
dimension $2n(4n + 1)$.  This raises the possibility of a hierarchy of theories, where the multiparticle states of one level, $k$, say, constitute the
single-particle states at level $2k$.  In this sense the Zhang-Hu four-dimensional theory could be construed as the theory of quantum ensembles of
two dimensional theories of the quantum Hall effect. 
\eject
\subsection*{Complex twistor geometry; the Plucker quadric $\Q$}Let $\V$ be a vector space over a field $\F$, of finite dimension $n$.  We
associate to
$\V$ the following entities:
\begin{itemize}\item $\V^*$ the dual space of $\V$. 
\item $\GL(\V)$ the group of all isomorphisms of $\V$.
\item $\Gr(k, \V)$ the Grassmanian of all
$k$-dimensional subspaces of $\V$, where $k$ is an integer in the range $0 \le k \le n$.
\item  $\PV$ the projective space of $\V$: $\PV = \Gr(1, \V)$. 
\item For $\X\subset \V$, we put $\Gr(k, \X) = \{ x\in \Gr(k, \V); x\cap (\X -\{0\})\ne \phi\}$; we abbreviate $\Gr(1, \X)$ by $\PX$,
called the projective image of $\X$; also, for $x \in \V - \{0\}$, we abbreviate $\mathcal{P}\{x\}$ by $\mathcal{P}x$.
\item $\Omega^k(\V)$, the $k$-th  exterior power of $\V$, for any non-negative integer $k$, where we take $\Omega^0(\V) = \F$ and   $\Omega^1(\V) =
\V$.
\item The map $\omega^k_{\V}:\Gr(k, \V)\rightarrow \textrm{P}\Omega^k(\V)$, which for any $x \in \Gr(k, \V)$ with basis $\{v_i: i = 1,2, \dots,
k\}$, maps $x$ to the element of $\mathcal{P}\Omega^k(\V)$ given as the projective image of the exterior product $\Pi_{i =1}^k
v_i$.
\item The spaces $\Gr(k, \V^*)$, which may be identified with $\Gr(n - k, \V)$.
\end{itemize}
For this work, we study twistor space, which may be considered to be a complex vector space
$\T$ of complex dimension four.  Its natural symmetry group
$\GL(\T)$ has complex dimension sixteen. We associate to $\T$ the following: \begin{itemize}\item The affine Plucker quadric, $\AQ$, is
the subset of
$\Omega^2(\T)$ consisting of all $X\in \Omega^2(T)$, such that $X\wedge X = 0$. $\AQ$ is a five dimensional submanifold $\Omega^2(\T)$, singular
only at the origin.
\item  The Plucker quadric $\Q$, is the projective image of $\AQ$ in $\mathcal{P}\Omega^2(\T)$.  Then $\Q$ is a holomorphic four
dimensional submanifold of $\mathcal{P}\Omega^2(\T)$ and an orbit for $\GL(\T)$.
\item The map $\omega^2_{\T}: \Gr(2, \T)\rightarrow \mathcal{P}\Omega^2(\T)$ has image
$\Q$ and gives a natural isomorphism of $\Gr(2, \T)$ with $\Q$, equivariant with respect to the natural actions of $\GL(\T)$. 
\item The space $\Q$ has a natural holomorphic, conformally flat,
conformal structure, such that, for $x$ and $y$ in $\Gr(2, \T)$, the points $\omega_{\T}^2(x)$ and $\omega^2_{\T}(y)$ in $\Q$ are null related
(connected by a null geodesic), iff
$x + y \ne \T$.  
\item In general, there is no natural metric representing the conformal structure
of $\Q$.  However, if a symplectic form $S$ is given on $\T$, then the tensor $g_s$, given by the formula: $(S\wedge S).(dX \wedge dX) =
(S.X)^2g_S$, defined for $X\in \Omega^2(T)$, with $S.X\ne 0$, induces, on restriction to
$\AQ$, a canonical representative of the conformal structure of
$\Q$, also called $g_S$,  defined globally, except on the non-null hypersurface $\Q_S$, with equation $S.X = 0$.  Then $g_S$ gives $\Q - \Q_S$ a
metric of constant non-zero curvature.
\item  The natural symmetry group preserving the metric $g_S$ is now the subgroup of
$\GL(\T)$, preserving
$S$.  This is a complex symplectic group of complex dimension ten.  It acts on $\Q - \Q_S$ as the complex orthogonal group $\SO(5,
\C)$.
\end{itemize}\subsection*{The Euclidean reality structure; the sphere $\Q_E$; the symmetry reduction from $\GL(\T, \HH)$ to
$\Spin(5, \R)$}  Given a twistor space $\T$, a Euclidean reality structure is a quaternionic conjugation on $\T$:  a complex antilinear, real
linear isomorphism of
$\T$ with square minus the identity.  The conjugation gives $\T$ the structure of a two-dimensional quaternionic vector space.  For a tensor $\tau$
based on
$\T$, denote its conjugate by
$\overline{\tau}$. 
\begin{itemize}\item Then the are no non-zero real
twistors, but there is a real four dimensional part of the Plucker quadric, denoted $\Q_E$, topologically a four-sphere.  This is the four-sphere
of the Zhang-Hu theory.
\item   The symmetry group is now
$\GL(\T, \HH)$ the Lie group of all isomorphisms of
$\T$ preserving the quaternionic conjugation.  It has real dimension 16 and is isomorphic to the Lie group of all invertible $2\times 2$
matrices of quaternions.  It contains $\Spin(1,5,\R)$ as a normal subgroup of real co-dimension one. 
\item If now $\T$ is equipped also with a
symplectic structure
$S$, that is real and positive definite with respect to the quaternionic conjugation (so $S = \overline{S}$ and $S(z, \overline{z}) > 0$, for any
non-zero
$z
\in
\T$), then the symmetry group reduces to the quaternionic unitary group $\U(2, \HH)$, which is isomorphic to the group $\Spin(5,\R)$.  In this case
the natural metric
$g_S$ of the Plucker quadric, associated to $S$, gives the real four-sphere its standard (real) round metric.
\end{itemize}
\subsection*{The Lorentzian reality structure; the hyperquadric; the spacetime $\Q_L$; the action of $\SO(4, \R)$} 
Given a twistor space $\T$, a Lorentzian reality structure is a conjugation mapping $\T$ to $\T^*$ and vice-versa, such that the conjugate of the
conjugate of a twistor gives back the original twistor and such that the conjugation has signature $(2,2)$: if $Z\in\T$, with
conjugate
$Z'\in\T^*$, then the hermitian form $Z'.Z$ can be diagonalized as $|a|^2 + |b|^2 - |c|^2 - |d|^2$, where $(a,b,c,d) \in \C^4$ are the co-ordinates
of
$Z$, with respect to a suitable basis for $\T$.  For a tensor $\tau$ based on $\T$, denote its Lorentzian conjugate by
$\tau'$.  The twistor space $\T$ separates into three spaces, $\T^{\pm}$ and $\N$ and the projective space $\PT$ into the
corresponding projective images, $\PT^{\pm}$ and $\PN$:
\begin{itemize}\item $\T^{\pm}$ consists of all twistors for which $\pm Z'.Z > 0$.  The spaces $\T^{\pm}$ and $\PT^{\pm}$ are open complex
submanifolds of $\T$ and $\PT$, respectively. 
\item $\N$ consists of all null twistors: those that obey $Z'.Z = 0$.  $\N$ is closed and is the common boundary of $\T^\pm$.  Away from the
origin, $\N$ is a smooth real hypersurface and has a Levi-form of signature $(+, -, 0)$. 
\item $\PN$ is the compact
hyperquadric of indefinite Levi-form, with flat Chern-Moser-Webster connection.  The real analytic manifold $\PN$ has real dimension five
and has topology
$S^3\times S^2$.  $\PN$ is closed in $\PT$ and is the common boundary of $\PT^{\pm}$. Its Fefferman conformal structure is defined on $S^3\times S^3$ and
may be taken to be $\pi_1^*(g) - \pi_2^*(g)$, where $g$ is the round metric of $S^3$ and $\pi_1: S^3\times S^3\rightarrow S^3$ and $\pi_2: S^3\times
S^3\rightarrow S^3$ are the natural projections given by the formulas: $\pi_1(x,y) = \pi_2(y, x) = x$, for any $(x, y) \in S^3\times S^3$.
\end{itemize}  
A point $\omega^2_{\T}(x)$ of the Plucker quadric (with $x \in \Gr(2, \T)$) is said to be real iff the twistor hermitian form
vanishes on restriction to
$x$, iff the plane $x$ lies in $\N$.  
\begin{itemize} \item There is a four dimensional family
$\Q_L$ of real points, and the complex conformal structure of the Plucker quadric induces a real Lorentzian conformal structure on $\Q_L$, which is
then a conformal compactification of real Minkowski space-time. 
\item  The space $\Q_L$ may be realized concretely as the group $\U(2, \C)$, with the
conformally flat Lorentzian metric $g = -\frac{1}{4}\det(\theta)$, where $\theta$ is the Maurer-Cartan form of $\U(2, \C)$, regarded as an
anti-hermitian
$2$ by $2$ matrix of one-forms. $g$ is both left and right invariant under the action of $\U(2, \C)$ on itself. 
\item The symmetry group is now
$\U(2,2, \C)$, the group of all isomorphisms of
$\T$ preserving the hermitian form $Z'.Z$.  It acts on $\Q_L$ as the conformal group. 
\item  In terms of the Plucker quadric, the projective null twistors
$\mathcal{P}Z$, for $Z \ne 0 $ in $\N$ are represented by the elements $X \ne 0$ in $\AQ$, representing real points, such that $X\wedge Z =
0$.  For a fixed
$Z$, these points form a topological circle and constitute a null geodesic on
$\Q_L$. This gives a natural isomorphism between $\PN$ and the space of all null geodesics in $\Q_L$.
\item In the language of $\U(2, \C)$, a null geodesic
through the identity is the one parameter subgroup
$[N] =
\{
\exp(itN); t
\in
\R\}$ where
$N$ is  a rank-one hermitian matrix (the null geodesics through other points are obtained from those passing through the identity by left
translation).   The one-parameter subgroup
$[N]$ is periodic in the parameter
$t$, with period
$\frac{2\pi}{\textrm{tr}(N)}$.  Then the correspondence between 
$\PN$ and the space of null geodesics of $\Q_L$ may be described explicitly as follows.
\begin{itemize} \item The points of $\N$ may be described by pairs $(\alpha, \beta)$
where
$\alpha$ and $\beta$ are elements of $\C^2$, such that $ |\alpha|^2 = |\beta|^2$.
\item Then each point of $\PN$ is the projective image of an $(\alpha, \beta) \in \N$, where both $\alpha$ and $\beta$ are non-zero. 
\item   The null geodesic
corresponding to
$(\alpha, \beta)$ is the set of all
$X\in \U(2,\C)$ such that
$\det(X)\alpha = X\beta$. 
\end{itemize}
\item  When $\Q_L$ is represented as $\U(2, \C)$, it carries an isometric action of
$\SO(4,\R)$, where $\SO(4, \R)$ is regarded as the quotient of
$\SU(2,
\C)\times\SU(2,
\C)$ by the identification of each
$(A, B)
\in \SU(2,
\C)\times \SU(2,\C)$ with $(-A, -B)$: the action is
$X\rightarrow AXB^{-1}$, defined for any $X\in \U(2, \C)$, where
$A$ and
$B$ lie in $\SU(2, \C)$: the pair $(A,B)$ and $(-A,-B)$ give the same action, so the action passes down the quotient group $\SO(4,
\R)$.
\item Note that the $\SO(4,
\R)$ action on $\U(2, \C)$ preserves $\det(X)$, so the spacelike hypersurfaces $\det(X) = \textrm{constant}$ are $\SO(4, \R)$ invariant. 
\item The induced action on $\PN$ corresponding to the pair $(A, B)$ is $(\alpha, \beta) \rightarrow (A\alpha, B\beta)$.   
\item Finally note
that the points in
$\U(2,
\C)$ along any null geodesic take each possible value of $\det(X)$ exactly once, so the space $\PN$ is fibred over every surface $\det(X) =
\textrm{constant}$. 
\end{itemize}   
\subsection*{The twistor fibration}
When the twistor space $\T$ has a quaternionic conjugation, the space $\T-\{0\}$ fibres naturally over the real part of $\AQ$
and over $\Q_E$, the latter being the Euclidean real part of the Plucker quadric.  The natural projections map $Z\in \T - \{0\}$ to:
\begin{itemize} \item $Z\wedge \overline{Z}$ in $\AQ$ ; the fibre is a three-sphere: the ensemble of all linear combinations $\rho Z + \sigma
\overline{Z}$ with $(\rho, \sigma) \in \C^2$ and
$|\rho|^2 + |\sigma|^2 = 1$.
\item $\textrm{Span}(Z, \overline{Z})$ in $\Gr(2,T)$; the fibre is $\C^2 - \{0\}$: the ensemble of all linear combinations $\rho Z +
\sigma
\overline{Z}$ with $(\rho,\sigma) \in \C^2 - \{0\}$.
\item $\mathcal{P}(Z\wedge \overline{Z})$ in the Plucker quadric; the fibre is $\C^2 - \{0\}$: the ensemble of all linear combinations $\rho Z
+
\sigma\overline{Z}$ with $(\rho, \sigma) \in \C^2 - \{0\}$.
\end{itemize}
This fibration is the key to the Zhang-Hu approach: a holomorphic function of $Z$ is thought of as a generalized "function" on the four-sphere, where
the point of the four-sphere $\Q_E$ associated to $Z$ is precisely $\mathcal{P}(Z\wedge \overline{Z})$.
\subsection*{Combining the reality structures}The group $\U(2,2,\C)$ maps epimorphically to the group $\SO(2,4, \R)$.  The group $\GL(2, \HH)$ maps
epimorphically to $\SO(5, \R)$.  The two groups $\SO(2,4, \R)$ and $\SO(5, \R)$ clearly have the group $\SO(4, \R)$ in common.  So it is possible,
as seen by Zhang and Hu, that breaking the symmetry of the Zhang-Hu theory down to $\SO(4, \R)$ allows one to connect with relativity.  This may be
done by choosing a stratification of the four-sphere by level surfaces of a function invariant under $\SO(4, \R)$, but not the full group $\SO(5,
\R)$.   Geometrically, this may be achieved by using the Lorentzian and Euclidean reality structures of $\T$ simultaneously.\\
\\
If $Z\rightarrow \overline{Z}$ and $Z\rightarrow Z'$ are the Lorentzian and Euclidean reality structures, respectively, then we require:
\begin{itemize}\item  The two structures should commute: $\overline{Z'} = (\overline{Z})'$.     
\item The complex-linear map from $\T $ to $\T^*$, $Z \in \T \rightarrow (\overline{Z})'\in \T^*$ should be symplectic.  
\end{itemize} 
\eject\noindent If now the twistor is null with respect to the Lorentzian structure: $Z'.Z = 0$, then the entire plane spanned by 
$Z$ and $
\overline{Z}$ (with $Z\ne 0$) is also null, so it is real with respect to both structures.  The signature of the hermitian form $Z'.Z$
when restricted to the plane through $Z$ and $\overline{Z}$ is one of only three types (positive definite, negative definite, or identically
zero).  Projectively, the positive definite planes foliate $\PT^+$, the negative definite planes foliate $\PT^-$ and the planes with trivial
hermitian form, foliate $\PN$.  These then give a preferred three real dimensional space-like hypersurface in $\Q_L$ and constitute the
intersection of
$\Q_L$ and
$\Q_E$ in $\Q$.   \\
\\
From the point of view of Minkowski spacetime, there are two slightly different ways of regarding these structures, depending on where one places
the vertex of the null cone at infinity. To explain these, we will use the usual description of twistors by means of pairs of Lorentzian spinors. 
\\
\\Recall that there are two spin spaces, $\mathcal{S}$ and $\mathcal{S}'$ (called the unprimed and primed spinor spaces, respectively), each vector
spaces of two complex dimensions, each equipped with symplectic forms, denoted
$\epsilon$ and
$\epsilon'$, respectively, such that complex conjugation interchanges $\mathcal{S}$ and $\mathcal{S}'$ and $\epsilon$ and $\epsilon'$.  They carry
the fundamental representations of the group
$\SL(2,
\C)$.  We also need the dual spaces $\mathcal{S}^*$ and $(\mathcal{S}')^*$ called the unprimed and primed co-spin spaces, respectively.  These
are equipped with the inverse symplectic forms $\epsilon^{-1}$ and $(\epsilon')^{-1}$, respectively.   The real
vector representation of
$\SL(2,
\C)$ is the self-conjugate part of
$\mathcal{S}\otimes_{\C}\mathcal{S}'$, for which
$ g= \epsilon\otimes_{\C}\epsilon'$ provides a real Lorentzian metric.  Using indices, an element of $\mathcal{S}$ would be denoted $\omega^A$, its
conjugate in $\mathcal{S}'$ by $\overline{\omega}^{A'}$ and a (real or complex) vector by $t^{AA'}$, which is abbreviated $t^a$.  Indices are
raised and lowered using the symplectic structures $\epsilon_{AB}$ and $\epsilon'_{A'B'}$  and their inverses  $\epsilon^{AB}$ and
$(\epsilon')^{A'B'}$  according to the scheme: given $v^A$, then $v_A = v^B\epsilon_{BA}$ and $v^A = \epsilon^{AB}v_B$, together with their
conjugates.\\  \\  Using spinors, a twistor
$Z$ and a dual twistor $W$ can be described as spinor pairs:
\begin{itemize}\item A twistor: $Z = (\omega,
\pi)$, consisting of elements $\omega\in \mathcal{S}$ and $\pi \in (\mathcal{S}')^*$.
\item A dual twistor: $W = (\alpha, \beta)$ consisting of elements $ \alpha\in \mathcal{S}^*$
and
$
\beta
\in \mathcal{S}' $.
\item The dual pairing is then: $W.Z = \alpha.\omega + \pi.\beta$. 
\item Using indices, we write:
$Z^\alpha = (\omega^A,
\pi_{A'})$, $W_\alpha = (\alpha_A, \beta^{A'})$ and $W_\alpha Z^\alpha = \alpha_A\omega^A + \beta^{A'}\pi_{A'}$. 
\item  The Lorentzian conjugation
is standard:
$Z'_\alpha = (\overline{\pi}_A,
\overline{\omega}^{A'})$, $(W')^\alpha = (\overline{\beta}^A, \overline{\alpha}_{A'})$.
\end{itemize}
\eject\noindent
The two representations of the Euclidean conjugation are then as follows:
\begin{itemize}\item  For the first, we take the Euclidean conjugation to be given by the formula: $\overline{Z}^\alpha =
(\overline{\omega}^{B'}t_{B'}^A, -\overline{\pi}_Bt_{A'}^B)$ and $\overline{W}_\alpha =
(-\overline{\alpha}_{B'}t^{B'}_A, \overline{\beta}^{B}t_{B}^{A'})$, where
$t^a$ is a real future pointing timelike vector of Lorentzian squared length two: $t^at_a = 2$.  Then we have $(\overline{Z})'_\alpha =
(-
\pi_{B'}t_{A}^B, \omega^{B}t_B^{A'}) = Z^{\beta}T_{\alpha\beta}$, where the components of $T_{\alpha\beta}$ are as follows: $T_{AB} = 0$,
$T^{A'B'} = 0$,
$T_A^{\hspace{4pt} B'} = -t_A^{B'}$ and
$T^{A'}_{\hspace{5pt}B} = t_B^{A'}$.  Evidently $T_{\alpha\beta}$ is self-conjugate, skew and symplectic, as required and the conjugations obey the
required compatibility conditions. 
\item For the second representation, we take the Euclidean conjugation to be given by the formula: $\overline{Z}^\alpha = (\overline{\pi}^{A}, -
\overline{\omega}_{A'})$ and $\overline{W}_\alpha = (-\overline{\beta}_A, \overline{\alpha}^{A'})$.   Then we have $(\overline{Z})'_\alpha =
(-\omega_A, \pi^{A'}) = Z^{\beta}U_{\alpha\beta}$, where the components of $U_{\alpha\beta}$ are as follows: $U_{AB} = \epsilon_{AB}$,
$U^{A'B'} = \epsilon^{A'B'}$,
$U_A^{\hspace{3pt} B'} = 0$ and
$U^{A'}_{\hspace{3pt}B} = 0$.  Evidently $U_{\alpha\beta}$ is self-conjugate, skew and symplectic, as required and the conjugations obey the
required compatibility conditions. 
\end{itemize}
The positive definite hermitian form on $\T$ is given in both cases by the formula:
\begin{equation}\omega(Z, \overline{Z}) = t_{a}\omega^A\overline{\omega}^{A'} + t^{a}\pi_{A'}\overline{\pi}_{A'}.\end{equation}
\begin{itemize}\item In the first case, the corresponding symplectic form, such that $\omega(Z, \overline{Z}) =
\omega_{\alpha\beta}Z^\alpha\overline{Z}^\beta$ is given by the formula $\omega_{\alpha\beta} = -U_{\alpha\beta}$. \item  In the second case, the
symplectic form is $\omega_{\alpha\beta} = -T_{\alpha\beta}$.\end{itemize}
Turning to the quantum commutation relations, written using indices, these are:
\begin{equation} [Z^\alpha, Z^\beta] = [\overline{Z}^\alpha, \overline{Z}^\beta] = 0, \hspace{10pt}[Z^\alpha, \overline{Z}^\beta] =
-(\omega^{-1})^{\alpha\beta}.\end{equation}
Writing these out in terms of the basic spinors $\omega^A$ and $\pi_{A'}$, the relations turn out to be the same for each of the two cases.  
The non-zero commutation relations are:
\begin{equation} [\omega^A, \overline{\omega}^{A'}] = - t^a, \hspace{10pt} [\pi_{A'}, \overline{\pi}_{A}] = - t_a.\end{equation}  
Quantization has $\omega^A$ and $\pi_{A'}$ as multiplication operators on entire functions of $Z^\alpha$, whereas the quantities
$\overline{\omega}^{A'}$ and
$\overline{\pi}_A$ are quantized as derivative operators:
\begin{equation} \overline{\omega}^{A'} = t^a\frac{\partial}{\partial \omega^A},  \hspace{20pt}\overline{\pi}_{A} =
t_a\frac{\partial}{\partial \pi_{A'}}.\end{equation} The twistor norm $N = Z'.Z = \overline{\pi}_A\omega^A +
\overline{\omega}^{A'}\pi_{A'}$ has the unambiguous quantization:
\begin{equation}N = Z'.Z = \omega^At_a\frac{\partial}{\partial \pi_{A'}} + \pi_{A'}t^a\frac{\partial}{\partial \omega^A}.\end{equation}
This formula is most easily understood by introducing new variables: 
\begin{equation} \alpha^A = \omega^A + t^a\pi_{A'}, \hspace{20pt}\beta^A = \omega^A - t^a\pi_{A'}, \end{equation}\begin{equation}\omega^A =
\frac{1}{2}(\alpha^A +
\beta^A),
\hspace{20pt}\pi_{A'} = \frac{1}{2}t_a(\alpha^A - \beta^A). 
\end{equation}
Then we have the two inner product formulas:
\begin{equation} Z'.Z = \frac{1}{2}t_a(\alpha^A\overline{\alpha}^{A'} - \beta^A\overline{\beta}^{A'}), \hspace{20pt}  \omega(Z,
\overline{Z}) =
\frac{1}{2}t_a(\alpha^A\overline{\alpha}^{A'} + \beta^A\overline{\beta}^{A'}). \end{equation}
The twistor is now represented by the spinor pair $Z^\alpha = (\alpha^A, \beta^A)$.  The Lorentzian conjugate of $Z^\alpha$ is now $Z'_\alpha =
\frac{1}{2}(t_a\overline{\alpha}^{A'}, - t_a\overline{\beta}^{A'})$.  The Euclidean conjugate of $Z^\alpha$ is $\overline{Z}^\alpha =
(t_{A'}^A\overline{\alpha}^{A'}, t_{A'}^A\overline{\beta}^{A'}) $ in the first case and  $\overline{Z}^\alpha =
(t_{A'}^A\overline{\alpha}^{A'}, -t_{A'}^A\overline{\beta}^{A'}) $ in the second case.  Then the operator
$N$ acts simply on the spinors
$\alpha^A$ and
$\beta^A$:
\begin{equation} N(\alpha^A) = \alpha^A, \hspace{20pt} N(\beta^A) =  - \beta^A,\end{equation}\begin{equation} N =
\alpha^A\frac{\partial}{\partial \alpha^A} - \beta^A\frac{\partial}{\partial \beta^A}.\end{equation}In the $\U(2, \C)$ formalism,
discussed above, the spinors
$\alpha^A$ and $\beta^A$ carry the fundamental representations of the product group
$\SU(2,
\C)\times
\SU(2,
\C)$, one trivial on the first factor of the product, the other trivial on the second factor. \\
\\
Finally the quantum commutation relations are:
\begin{equation} [\alpha^A, \overline{\alpha}^{A'}] = - 2t^a,\hspace{10pt} [\alpha^A, \overline{\beta}^{A'}] = 0,\hspace{10pt} [\beta^A,
\overline{\beta}^{A'}] = - 2t^a. \end{equation}
The other commutators: of $\alpha^A$ with $\alpha^B$, of $\alpha^A$ with $\beta^B$ and of $\beta^A$ with $\beta^B$ and all their conjugates, all
are zero.  These commutation relations are manifestly invariant under the action of the group $\SU(2,
\C)\times
\SU(2,
\C)$.

\eject
\subsection*{The abstract formulation of the particle-hole operators for the edge-states} Consider a Grassman variable $\theta$ ranging over a finite
dimensional Hilbert space.  The number operator giving the number of particles in a state is $N = \theta.\delta_{\theta}$.  If now the identity
operator $I$ is broken up into
$m$ pieces:
$I = \sum_{k = 1}^m J_k$, where each $J_k$, for $k = 1, 2, \dots, m$ is a projection operator, then the number operator is similarly broken up:
\begin{equation}N =
\sum_{k = 1}^m N_k, \hspace{10pt} N_k = \theta.J_k.\delta_\theta, \hspace{10pt} k = 1,2,\dots, m.\end{equation}
The application here is to the Zhang-Hu Hilbert space, which at level $r$ (for us $r$ is odd) has dimension $\binom{r + 2n - 1}{2n-1}$. This
forms an irreducible representation of $\U(2, \HH)$.  When we break the symmetry down to $\SO(4, \R)$, the Hilbert space will decompose into
irreducible representations of the smaller group.  There is then a corresponding decomposition of the number operator. Putting together two such pieces
gives the  particle-hole operator described by Zhang and Hu.  Here the required breaking of the symmetry can be achieved at the classical level by
foliating the four-sphere into three-spheres, using the level surfaces of the classical function $Z'.Z$ (or projectively the function
$\frac{Z'.Z}{\omega(Z,
\overline{Z})}$),  precisely the function used in ordinary twistor theory to define the null twistors.  However the Zhang-Hu theory requires that this be
done at the quantum level.  Then we need the quantum operator corresponding to
$Z'.Z$.  But, using the variables of the last section, this is just the operator
$N = \alpha^A\frac{\partial}{\partial \alpha^A} - \beta^A\frac{\partial}{\partial \beta^A}$. At level $r$ the states are already homogeneous of
degree $r$ in the twistor $Z^\alpha = (\alpha^A, \beta^A)$, so are eigenstates, with eigenvalue
$r$ of the operator $H = \alpha^A\frac{\partial}{\partial \alpha^A} + \beta^A\frac{\partial}{\partial \beta^A}$.  Thus the edge states are
characterized by states that are simultaneously eigenstates of the operators $H$ and $N$, or equivalently, simultaneously eigenstates of the
homogeneity operators $A = \alpha^A\frac{\partial}{\partial \alpha^A}$ and $B =  \beta^A\frac{\partial}{\partial \beta^A}$.  So they are
separately homogenous in the variables $\alpha^A$ and $\beta^A$: thus they form an irreducible representation of the group $\SU(2, \C)\times
\SU(2, \C)$, in agreement with the Zhang-Hu theory. \\\\ So at level $r$, the edge states will employ polynomials $f_{p,q}(\alpha^A,
\beta^A)$, homogeneous of degree $p$ in $\alpha^A$ and $q$ in $\beta^A$, where
$p$ and $q$ are non-negative integers such that $p + q = r$.   These form a representation of the Lie group $\SU(2, \C)\times
\SU(2, \C)$ of complex dimension $(p+1)(q+1)$.  
\eject\noindent Next we write the Grassman
variable at level
$r$ as
$\theta^{\alpha_1\alpha_2\dots \alpha_r}$, which is symmetric its indices, with the corresponding derivation
$\delta_{\beta_1\beta_2\dots \beta_r}$ (also symmetric), such that $\delta_{\beta_1\beta_2\dots \beta_r}\theta^{\alpha_1\alpha_2\dots \alpha_r}  =
\delta_{(\beta_1}^{(\alpha_1}\delta_{\beta_2}^{\alpha_2}\dots \delta_{\beta_r)}^{\alpha_r)}$.  Then the splitting of the twistor space $\T$ into
the direct sum $\T = \mathcal{S} \oplus \mathcal{S}$, corresponding to writing the twistor $Z^\alpha$ as the pair $(\alpha^A, \beta^A)$ gives a
splitting of the Kronecker delta tensor
$\delta_\alpha^\beta = P_\alpha^\beta + Q_\alpha^\beta$, where
$P_\alpha^\beta$ acts as the identity on the first term in the direct sum and kills the second term, whereas
$Q_\alpha^\beta$ kills the first term and acts as the identity on the second.  In turn, this splitting gives the splitting of the identity operator
$ \delta_{(\beta_1}^{(\alpha_1}\delta_{\beta_2}^{\alpha_2}\dots \delta_{\beta_r)}^{\alpha_r)}$ for the representation
$\T^r$:
\begin{equation}\delta_{(\beta_1}^{(\alpha_1}\delta_{\beta_2}^{\alpha_2}\dots \delta_{\beta_r)}^{\alpha_r)} = \sum_{p + q = r}
(J_{p,q})_{\beta_1\beta_2\dots \beta_r}^{\alpha_1\alpha_2\dots \alpha_r}, \end{equation}\begin{equation} (J_{p,q})_{\beta_1\beta_2\dots
\beta_{p+q}}^{\alpha_1\alpha_2\dots \alpha_{p+q}} = \frac{(p+q)!}{p!q!} P_{(\beta_1}^{(\alpha_1}P_{\beta_2}^{\alpha_2}\dots
P_{\beta_p}^{\alpha_p}Q_{\beta_{p+1}}^{\alpha_{p+1}} Q_{\beta_{p+2}}^{\alpha_{p+2}}
\dots Q_{\beta_{p+q})}^{\alpha_{p+q})}.\end{equation}
For convenience, we also write these expressions in index-free language.  For any $Z = (\alpha,
\beta)
\in
\T$ and
$W = (\rho, \sigma)\in \T^*$, with
$\alpha\in \mathcal{S}$, $\beta
\in
\mathcal{S}$,  $\rho \in \mathcal{S}^*$ and $\sigma
\in
\mathcal{S}^*$, we have:
\begin{equation} (Z^r.\delta_r)(\theta_r.W^r) = (W.Z)^r = (\rho.\alpha + \sigma.\beta)^r, \end{equation}\begin{equation} PZ = (\alpha, 0),
\hspace{10pt}WP = (\rho, 0), \hspace{10pt}QZ = (0, \beta), \hspace{10pt}WQ = (0, \sigma), \end{equation}\begin{equation} WPZ = \rho.\alpha, \hspace{10pt}
WQZ = \sigma.\beta, \end{equation}\begin{equation} W^{p+q}J_{p, q}Z^{p+q} = \frac{(p+q)!}{p!q!}(WPZ)^p(WQZ)^q.\end{equation}
Then the Zhang-Hu edge operator $ E_{p,q,p'q'}(Z, \overline{Z})$ is a derivation of degree zero of the Grassmann algebra that is \emph{local} in twistor
space, in the sense that it depends on the choice of a twistor $Z$. Here
$p$, $q$, $p'$ and $q'$ are non-negative integers such that $p + q = p' + q' = r$.  Explicitly, the edge operator is given by the formula:
\begin{equation}E_{p,q,p'q'}(Z,
\overline{Z}) =
\theta.J_{p',q'}.E(Z, \overline{Z})^r.J_{p,q}.\delta.\end{equation}
Here $ E(Z, \overline{Z})^r$ is the $r$-fold totally symmetrized tensor product of the tensor $ E(Z, \overline{Z})\in \T\otimes \T^*$ and, in turn, the
tensor $ E(Z,
\overline{Z})$ is given by the formula, valid for any twistor $T$:
\begin{equation} E(Z, \overline{Z}).T = Z\omega(\overline{Z}, T) - \overline{Z}\omega(Z, T). \end{equation}
Then the Hamiltonian $Z'.Z$ provides the evolution of the edge operator: explicitly if $Z = (\alpha, \beta)$, then the evolution $Z(t)$ with parameter $t$
is
$Z(t) = (e^{it}\alpha, e^{-it}\beta)$. 
\subsection*{Discussion: the twistor fluid, points and time asymmetry}
In the foregoing sections, we have recast and generalized the Zhang-Hu theory of the four dimensional quantum Hall effect.  All the main features
of their theory have been retained.  At the same time we have made some significant changes.
\begin{itemize}\item We have deployed the standard relation of spin to statistics, so the theory presented here would only be fermionic at level
$r$ if
$r$ is odd.
\item We have not had to use the basic 't Hooft-Polyakov Yang-Mills instanton at all.  The reason that we can do this successfully follows from a
remarkable property of that instanton, most easily seen by understanding it from the point of view of standard twistor theory.   Thanks to the
beautiful discovery of Ward, anti-self-dual Yang-Mills fields are encoded in twistor theory by means of suitable holomorphic vector bundles over
twistor space \cite{A31}. In the case of the basic 't Hooft-Polyakov instanton used by Zhang and Hu, it is well known that the twistor description of
this instanton is achieved by the so-called "null correlation bundle".\\
\\
Specifically we select a (constant) symplectic form $\omega$ in $\T$.  Then
for each projective twistor
$\mathcal{P}{Z}$, with $Z \in
\T - \{0\}$,  there is naturally associated a projective plane $Z^* = \omega(Z, )$ in $\PT$, which (since $\omega$ is skew) passes through the
point
$\mathcal{P}{Z}$ itself.  Then the space of all projective lines that pass through  $\mathcal{P}{Z}$  and lie in the plane $Z^*$ is a complex
projective line.  We have one such projective line for each projective twistor
$\mathcal{P}{Z}$.  This gives a bundle $\mathcal{P}\mathcal{B}$ over $\PT$, with fibre the complex projective line.  There is a naturally
associated vector bundle $\mathcal{B}$, with fibre $\C^2$ whose projective bundle is $\mathcal{P}\mathcal{B}$.  This vector bundle is the
called the null correlation bundle and by the Ward correspondence gives rise to the standard instanton in complexified Minkowski spacetime. 
Further, if appropriate reality conditions are applied (quaternionic conjugation on twistor space, etc.) then this instanton gives an everywhere
smooth
$\SU(2, \C)$ anti-self-dual solution of the Yang-Mills equations on the four-sphere.  The reality condition in particular requires that the
symplectic form be real with respect to the Euclidean conjugation of twistor space.  Then we have precisely the instanton used by Zhang and
Hu.   The point here is that the only information in this instanton field is the symplectic form itself and it is only the symplectic form that is
needed to develop the theory. 
\item  Having eliminated the Yang-Mills field, the remaining theory is purely geometrical.  There is no "isospin space" per se.  This is replaced
by two things: first the symplectic form used implicitly by Zhang and Hu for their Yang-Mills field is retained and gives the four-sphere (which a
priori has only a conformally flat structure) its round metric.  Secondly the twistors are fully geometrical: they are pairs of spinors, where the
spinors are spinors responsive to the conformal and metrical geometry.
\item The present geometrical reformulation of the Zhang-Hu theory has important consequences when we go beyond flat space.  The simplest way to
do this is to deform away from (suitable parts of) projective three-space.  Then the structure needed for the theory is precisely that needed for
solutions of the Euclidean vacuum Einstein equations with anti-self-dual Weyl curvature and with cosmological constant: equivalently these are
gravitational instantons: it was shown by Ward and Le Brun that the only structure needed for this in twistor space is a symplectic form on the
twistor space (the analogous theory without cosmological constant is essentially a limiting case and was developed earlier by Newman and
Penrose; it uses a pair of degenerate Poisson and degenerate symplectic structures), together with appropriate reality conditions \cite{A27}-\cite{A32}. 
So the arena here is that of
\emph{gravity} not (immediately) of Yang-Mills theory.
\item Perhaps the most striking aspect of the theory presented here is that it is a \emph{theory in twistor space}.  The spacetime used in the
theory is a derived consequence, not the fundamental ingredient.  This is definitely in accord with general twistor philosophy, which predicts
that indeed spacetime is a derived concept.  Here we may well have a quantum liquid, but the liquid is \emph{not} in spacetime, rather it is in
the twistor space itself.  Also the edge-states of Zhang and Hu are associated with the quantum analogue of the surface $Z'.Z = c\omega(Z,
\overline{Z})$, with
$c$ a constant.  This surface is a $\CR$ submanifold of $\T$, called a hyperquadric and gives a model of the space of null geodesics in
Minkowski spacetime.  Thus if we think of the edge-states as in some sense providing deformations of the boundary, they are associated with
deformations of the $\CR$-structure: thus with first cohomology and thus with massless particles, since it is well known that the first
$\CR$-cohomology groups in twistor space encapsulate the zero rest mass field equations, for arbitrary helicity. In fact one can go
further: the quantum liquid approach may explain \emph{why} it is precisely these groups which are so important.\eject\noindent
\item  Note that it is critical in the
Zhang-Hu theory (and here) that it is not just the classical surface $Z'.Z = c\omega(Z, \overline{Z})$ that is relevant, but the
corresponding quantum operator equation.  That this is so, is crystallized in their key formula: $px_5 = m_1 + m_2 -  m_3 -  m_4$ which determines
the latitude
$x_5$ of the edge in terms of integers associated with with the description of their states: so the latitude is quantized.  However the classical
hyperquadric is probably more relevant when certain thermodynamic limits are taken. 
\item This brings us, finally, to the question of the physical interpretation of the present theory.  This has to do with the nature and
consequences of the thermodyamic limits to be taken in the theory.  We have not yet analyzed these in detail: this is a major task for the future,
for which we have just laid out some of the ground rules.  We should note that the single edge in the style of Zhang and Hu is unlikely to be the
last word in the matter, just as it is not in the two-dimensional theory.  More likely is either a multiplicity of edges or a multiplicity of
liquids separated from other liquids: so we need to ask whether or not this is a multi-universe theory.  For the moment I would rather take the
more conservative approach that there is a single liquid, with a multitude (not necessarily countable) of edges.  Later one could
investigate the more speculative multi-universe scenario probably implicit in having a multiplicity of liquids.  Even within the single liquid
model, one might think that there is a multitude of universes, each associated with one of the edges.  However I prefer to focus on the following
working hypotheses:
\begin{itemize}\item Each edge is to be associated with a point of spacetime.
\item The quantum liquid glues spacetime together.
\item The Zhang-Hu theory is associated with an instant of time.  As time evolves, so does the liquid and the spacetime, hopefully in accordance
with Einstein's gravity, this being the non-linear theory to go with the linear theory of Zhang and Hu.
\item The evolution is naturally chiral and time asymmetric.  
\end{itemize}
The point at issue is the nature of the thermodynamic limit: there are two main possibilities, it would seem: first that the edge stays "large" in
twistor space, asymptotically becoming, say, the classical surface $Z'.Z = 0$ and second that the edge shrinks around a particular projective line
in the twistor  space: the analogous concept in the Zhang-Hu approach is that the edge shrinks around a preferred point of the
four-sphere, say the "north pole".\eject\noindent One should note here that hyperquadrics, which are flat in the sense of Chern-Moser, Webster and
Fefferman lie arbitrarily close to any given projective line in projective three-space.  Although ultimately both kinds of limit may play a role,
I prefer the scenario where the edge focusses down at a point.  Also there are many possible complicating factors, which need to be considered: for
example two or more "edges", each corresponding to the neighborhood of a line, may overlap.\\
\\The association with an instant of time seems natural from the way the theory is set up: it is essentially Euclidean and the intersection with
 a spacetime of Minkowskian type takes place on a spacelike hypersurface.\\
\\
That the theory should be chiral and time asymmetric seems natural in that, for all the edges, the quantum fluid is on one side of the edge and not
the other.  This is analogous in conventional twistor theory to preferring, say, $\PT^+$ to $\PT^-$;  but it is well known that one space
corresponds to positive energy, the other to negative and one refers to positive helicity, the other to negative. The arrow of time would
then be uniform, for all points in spacetime (for a single fluid).
\item Finally there is the question of the connection with the remainder of physics.  Recently the author conjectured a possible unification of
disparate ideas in physics and mathematics, particularly string theory and the Grothendieck theory of "dessins d'enfants", where the main structure
is that of the pseudo-Kahler twistor spaces associated to null hypersurfaces in spacetime, (principally the null
cones of points) \cite{A24}-\cite{A26}.  To make this unification possible seems to requires a certain accommodation on behalf of string theory:
particularly liberation from excessively special backgrounds and liberation from compactness in the "internal dimensions" and also a realization in
twistor theory that the twistor spaces in question may be devoid of the usual four parameter family of points, expected from the naive application of the
Kodaira-Newman-Penrose theory \cite{A27}-\cite{A29}.  The complete spacetime is then made up of a web of interconnecting such twistor spaces.  Since each
such space is intuitively associated with a particular point of spacetime, this approach would seem to be more or less compatible with the quantum
liquid picture, presented above, the liquid providing the framework, in some sense, for the web of twistor spaces and giving a mechanism for the
conjectured time and chiral asymmetry.
\end{itemize}


\newpage\begin{thebibliography}{30}
\bibitem{A1} J. Hu and S.C. Zhang, \hspace{3pt}\emph{Collective excitations at the boundary of a 4D quantum Hall
droplet},\hspace{3pt}Stanford University preprint, 2001.
\bibitem{A2} S.C. Zhang and J. Hu, \hspace{3pt}\emph{A Four-Dimensional Generalization of the Quantum Hall Effect},\hspace{3pt} Science
\textbf{294}(5543), 823-828, 2001.
\bibitem{A3} S.C. Zhang, \hspace{3pt}\emph{The Chern-Simons-Landau-Ginzburg theory of the fractional quantum Hall
effect},\hspace{3pt}International Journal of Modern Physics,
\textbf{6B}, 25-58, 1992. 
\bibitem{A4} R. Prange and S. Girvin,\hspace{3pt} \emph{The Quantum Hall Effect}, \hspace{3pt}Springer Verlag,
Berlin, Germany, 1990. 
\bibitem{A5} M. Stone,\hspace{3pt} \emph{Schur functions, chiral bosons and the quantum-Hall-effect edge states},\hspace{3pt} Physical Review
\textbf{42B}, 8399-8404, 1990.   
\bibitem{A6} R. Laughlin, \hspace{3pt}\emph{Anomalous Quantum Hall Effect: An Incompressible Quantum Fluid with Fractionally Charged
Excitaions},\hspace{3pt} Physical Review Letters
\textbf{50}, 1395-1398. 1983.   
\bibitem{A7} F.D.M. Haldane, \hspace{3pt}\emph{Fractional Quantization of the Hall Effect: A Hierarchy of Incompressible Quantum Fluid States},
\hspace{3pt}Physical Review Letters
\textbf{51}, 605-608, 1983. 
\bibitem{A8} C.N. Yang, \hspace{3pt}\emph{$\mathcal{SU}_2$ monopole harmonics}, \hspace{3pt}Journal of Mathematical Physics, \textbf{19},
2622-2627, 1978. 
\bibitem{A9} C.N. Yang,\hspace{3pt} \emph{Generalization of Dirac's monopole to $\mathcal{SU}_2$ gauge fields}, \hspace{3pt} Journal of
Mathematical Physics,
\textbf{19}, 320-328, 1978. 
\bibitem{A10} G. 't Hooft,\hspace{3pt} \emph{Computation of the quantum effects due to a four-dimensional pseudo-particle}, \hspace{3pt} Physical
Review
\textbf{14D}, 3432-3438, 1976. 
\bibitem{A11} R. Jackiw and C. Rebbi, \hspace{3pt}\emph{Conformal properties of a Yang-Mills pseudoparticle}, \hspace{3pt} Physical Review
\textbf{14D}, 517-523, 1976.  
\bibitem{A12} A. Belavin, A. Polyakov, A. Schwartz and Y. Tyupkin,\hspace{3pt} \emph{Pseudoparticle solutions of the Yang-Mills equations},
\hspace{3pt}Physics Letters
\textbf{59B}, 85-87, 1975.  
\bibitem{A121} R.E. Cutkosky, \hspace{3pt}\emph{Faddeev-Popov Zeros and Confinement of Color in a Hyperspherical Gauge Model}, Physical Review Letters,
\textbf{51}, 538-541, 1983.
\bibitem{A13}R. Gilmore, \hspace{3pt} \emph{Lie Groups Lie Algebras and
Some of Their Applications},  \hspace{3pt} John Wiley and Sons, New York, 1974.
\bibitem{A14} S.A. Huggett and K.P. Tod, \hspace{3pt}\emph{An Introduction to Twistor Theory, London Mathematical Society
Student Texts}, \textbf{4}, Second edition, Cambridge University Press, Cambridge, 1994.
\bibitem{A15}  R.S. Ward and R.O. Wells, Jr., \hspace{3pt}\emph{Twistor Geometry and Field Theory, Cambridge Monographs on Mathematical
Physics}, \hspace{3pt} Cambridge University Press, Cambridge, 1991.
\bibitem{A16} R. Penrose and W. Rindler,\hspace{3pt} \emph{ Spinors and space-time Volume 2: Spinor and twistor methods in space-time
geometry},
\hspace{3pt} Cambridge University Press, Cambridge, 1986.
\bibitem{A17} R. Penrose and W. Rindler,\hspace{3pt} \emph{ Spinors and space-time Volume 1: Two-spinor calculus and relativistic
fields},
\hspace{3pt} Cambridge University Press, Cambridge, 1984.
\bibitem{A18}G.A.J. Sparling, \hspace{3pt} \emph{Theory of massive particles. {I}. {A}lgebraic structure}, \hspace{3pt}Philosophical
Transactions of the Royal Society of London,
\textbf{A301}(1458), 27-74, 1981.
\bibitem{A185} M.F. Atiyah, N.J. Hitchin, I.M. Singert,\hspace{3pt} \emph{Self-duality in four dimensional Riemannian geometry}, \hspace{3pt}Proceedings
of the Royal Society of London,
\textbf{A362}, 425-461, 1978.
\bibitem{A19} R. Penrose and M.A.H. MacCallum, \hspace{3pt}\emph{Twistor theory: An Approach to the Quantization of Fields and Space-Time}, Physics
Letters \textbf{6C}, 214-316, 1972.
\bibitem{A20} R. Penrose, \hspace{3pt}\emph{Twistor algebra}, \hspace{3pt}Journal of Mathematical Physics, \textbf{8}, 345-366, 1967. 
\bibitem{A21} G.A.J. Sparling, \hspace{3pt}\emph{The
twistor theory of hypersurfaces in spacetime},\hspace{3pt} in \emph{Further
Advances in Twistor Theory,  Volume III: Curved Twistor Spaces}, \hspace{3pt}editors
L.J. Mason, L.P. Hughston, P.Z. Kobak and H. Brezis, CRC Press,
London, 2001.
\bibitem{A22} C.L. Fefferman,\hspace{3pt} \emph{Monge-Ampere
Equations, the Bergman Kernel, and the Geometry of Pseudoconvex
domains}, \hspace{3pt}Annals of Mathematics, Second Series, \textbf{103},
395-416, 1974.
\bibitem{A23} S.-S. Chern, J.K. Moser (with Appendix by S.M.
Webster),\hspace{3pt}
\emph{Real hypersurfaces in complex manifolds}, \hspace{3pt}Acta
Mathematica
\textbf{133} 219-271, 1974.
\bibitem{A24} D. Kapadia  and G.A.J. Sparling,\hspace{3pt}\emph{Glitch metrics}, \hspace{3pt}Laboratory of Axiomatics preprint,
2002.
\bibitem{A25} D. Kapadia  and G.A.J. Sparling, \hspace{3pt}\emph{A class of conformally Einstein metrics}, \hspace{3pt}Classical and Quantum
Gravity,
\textbf{24}, 4765-4776, 2000.
\bibitem{A26} G.A.J. Sparling,\hspace{3pt}
\emph{Zitterbewegung}, \hspace{3pt}Seminaires et Congres, Societe Mathematique de
France, \textbf{4}, 275-303, 2000.
\bibitem{A265}A. Grothendieck,  \hspace{3pt}\emph{Esquisse d'un programme} in \emph {Geometric Galois actions: Volume 1 Around Grothendieck's Esquisse
d'un programme}, \hspace{3pt} editors Leila Schneps and Pierre Lochak, London Mathematical Society Lecture Notes in Mathematcis \textbf{242}, Cambridge
University Press, Cambridge, 1997.
\bibitem{A266}M.B. Green, J.H. Schwarz, E. Witten, \hspace{3pt}\emph{Superstring Theory: Volume 1 Introduction}, Cambridge University Press, Cambridge,
1988.
\bibitem{A267}M.B. Green, J.H. Schwarz, E. Witten, \hspace{3pt}\emph{Superstring Theory: Volume 2 Amplitudes Anomalies and Phenomenology},
Cambridge University Press, Cambridge, 1988.
\bibitem{A27} K. Kodaira,\hspace{3pt} \emph{A theorem of completeness of characteristic systems
for analytic families of compact submanifolds of complex manifolds}, \hspace{3pt}Annals of Mathematics, \textbf{75}, 146-162, 1962.
\bibitem{A28} E.T. Newman, \hspace{3pt}\emph{Heaven and its properties},\hspace{3pt} General
Relativity and Gravitation, \textbf{7}, 107-127, 1976.
\bibitem{A29} R. Penrose,\hspace{3pt} \emph{Non-linear gravitons and curved twistor
theory},\hspace{3pt} General Relativity and Gravitation, \textbf{7}, 31-52, 1976.
\bibitem{A30} R.S. Ward,\hspace{3pt} \emph{Self-Dual Space-Times with Cosmological Constant}, \hspace{3pt}Communications in Mathematical
Physics,
\textbf{78}, 1-18, 1980.
\bibitem{A31} R.S. Ward, \hspace{3pt}\emph{On self-dual gauge fields},\hspace{3pt} Physics Letters, 
\textbf{61A}, 81-83, 1977.
\bibitem{A32} C.R. Le Brun,\hspace{3pt} \emph{$\mathcal{H}$-space with a cosmological constant}, \hspace{3pt}Proceedings of the Royal
Society of London,
\textbf{A380}, 171-185, 1982.
\bibitem{A33}G.A.J. Sparling, \hspace{3pt}\emph{The eth operator}, In \emph{Recent advances in general
relativity (Pittsburgh, Pennsylvania,
  1990)},  \emph{Essays in Honor of Ted Newman}, \emph{Einstein
Studies Volume 4}, \hspace{3pt}editors Allen I. Janis and John Porter,\hspace{3pt} Birkh\"auser
Boston, Boston, 26-59, 1991.


\end{thebibliography}
\end{document}